\documentclass[aps,prb,twocolumn,showpacs,floatfix,superscriptaddress,floatfix]{revtex4}
\usepackage[english]{babel}
\usepackage{graphicx}
\usepackage{color}
\usepackage[utf8]{inputenc}
\usepackage{pstricks,pst-grad,color}
\usepackage{graphicx,amssymb}
\usepackage{amsmath}
\usepackage{amssymb}

\begin{document}

\title{Magnetism in $f$ electron superlattices}


\author{Robert Peters}
\email[]{peters@scphys.kyoto-u.ac.jp}
\affiliation{Department of Physics, Kyoto University, Kyoto 606-8502,  Japan}

\author{Yasuhiro Tada}
\affiliation{Institute for Solid State Physics, The University of
  Tokyo, Kashiwa 277-8581, Japan}
\affiliation{Max Planck Institute for the Physics of Complex Systems, 01187 Dresden,
Germany}
\author{Norio Kawakami}
\affiliation{Department of Physics, Kyoto University, Kyoto 606-8502,  Japan}

\date{\today}


\begin{abstract}
We analyze antiferromagnetism in $f$ electron superlattices. We show that the competition between the Kondo effect and the RKKY interaction in $f$ electron materials is modified by the superlattice structure. Thus, the quantum critical point which separates the magnetic phase and the Fermi liquid phase depends on the structure of the $f$ electron superlattice.
The competition between the Kondo effect and the RKKY interaction is also reflected in the magnetic interlayer coupling between different $f$ electron layers. We demonstrate that in the case of weak Kondo effect the magnetic interlayer coupling behaves similar to other magnetic heterostructures without Kondo effect. However, close to the quantum phase transition, the dependence of the interlayer coupling on the distance between the $f$ electron layers is modified by the Kondo effect. Another remarkable effect, which is characteristic for $f$ electron superlattice, is that the magnetic interlayer coupling does vanish stepwise depending on the distance between different $f$ electron layers. As a consequence, the quantum critical point depends also stepwise on this distance.
\end{abstract}


\maketitle

\section{Introduction}

Recent experimental progress has made it possible to create thin layers of $f$ electron materials and artificial superlattices, which consist of a periodic structure of $f$ electron materials such as CeCoIn$_5$ and CeIn$_3$ and metals without $f$ electrons.\cite{Shishido2010,Mizukami2012,Goh2012,PhysRevLett.112.156404,PhysRevLett.116.206401,RPP_Shimozawa_2016}
Thus, it has become feasible to change the electronic structure and tune the properties of $f$ electron materials. This is particularly important when one thinks of the interesting phenomena which can be observed in these materials, such as magnetism, unconventional superconductivity and quantum criticality.
Usually, a quantum critical point in $f$ electron materials (if it exists) occurs at a certain pressure or magnetic field. These parameters are fixed by the electronic structure. In an artificially created $f$ electron superlattice, on the other hand, it is now possible to change the electronic structure of the material and thus tune the quantum critical point.
Furthermore, by combining layers of different $f$ electron materials the competition/interplay between quantum critical layers, superconducting layers, magnetic layers, metallic heavy-fermion layers can be studied, which opens invaluable opportunities to study novel phenomena.

An example of $f$ electron superlattices which has been recently created in laboratory  is CeIn$_3(n)$/LaIn$_3$(4). It has been experimentally shown that the N\'eel temperature in
CeIn$_3(n)$/LaIn$_3$(4) superlattices
decreases to zero when the Ce layer thickness $n$ is reduced to $n=2$,
which is accompanied by a linear temperature dependence of the
resistivity.\cite{Shishido2010} This demonstrates the influence of the superlattice structure on the magnetic state in the $f$ electron material and the ability to tune the quantum critical point.
In other experiments, using the heavy fermion
CeCoIn$_5$ and the conventional  metal YbCoIn$_5$, 
superconductivity has been observed in thin CeCoIn$_5$ layers. 
\cite{Mizukami2012}

The properties of $f$ electron materials are determined by the competition of the RKKY interaction and the Kondo effect. While the RKKY interaction favors a magnetically ordered state, the Kondo effect screens the magnetic moments of the strongly interacting $f$ electrons, which results in a paramagnetic Fermi liquid state. This competition is described and visualized in the Doniach phase diagram,\cite{doniach77}  which contrasts the energy scales of both effects; the RKKY interaction depends quadratically on the coupling between conduction electrons and $f$ electrons and the Kondo effect exponentially. Therefore, for small coupling the RKKY interaction is stronger than the Kondo effect, while for strong coupling the Kondo effect prevails.
When both effects are equal in strength, quantum criticality accompanied with non-Fermi liquid behavior can frequently be observed.\cite{coleman2007,coleman2005,gegenwart2008}

While the dependence of the RKKY interaction and the Kondo effect on the coupling between conduction electrons and $f$ electrons is well understood,\cite{doniach77} it is unclear how the competition of both depends on the superlattice structure.  
In this paper we analyze the influence of the superlattice structure on the RKKY interaction and the competition with the Kondo effect and, furthermore, study the dependence of the magnetic interlayer coupling on the distance between different $f$ electron layers and how it is modified by the Kondo effect.
We find that especially close to the quantum critical point, there are large modifications in the dependence on the distance between the $f$ electron layers from non-$f$ electron superlattices. The magnetic interlayer coupling vanishes stepwise when increasing the distance between different $f$ electron layers.
Thus, the quantum critical point changes stepwise when changing the number of spacer layers.

The remainder of this paper is organized as follows: In the next section we describe the model and methods which we use to analyze the $f$ electron superlattice. This is followed by section \ref{sec_Mag}, where we analyze the magnetization dependence in $f$ electron superlattices. 
Thereafter, we analyze the interlayer coupling in section \ref{sec_coupling} and the spectral functions in section \ref{sec_spectral}.
A conclusion finishes this paper.

\section{Model and Method}
\begin{figure}[t]
\begin{center}
  \includegraphics[width=\linewidth]{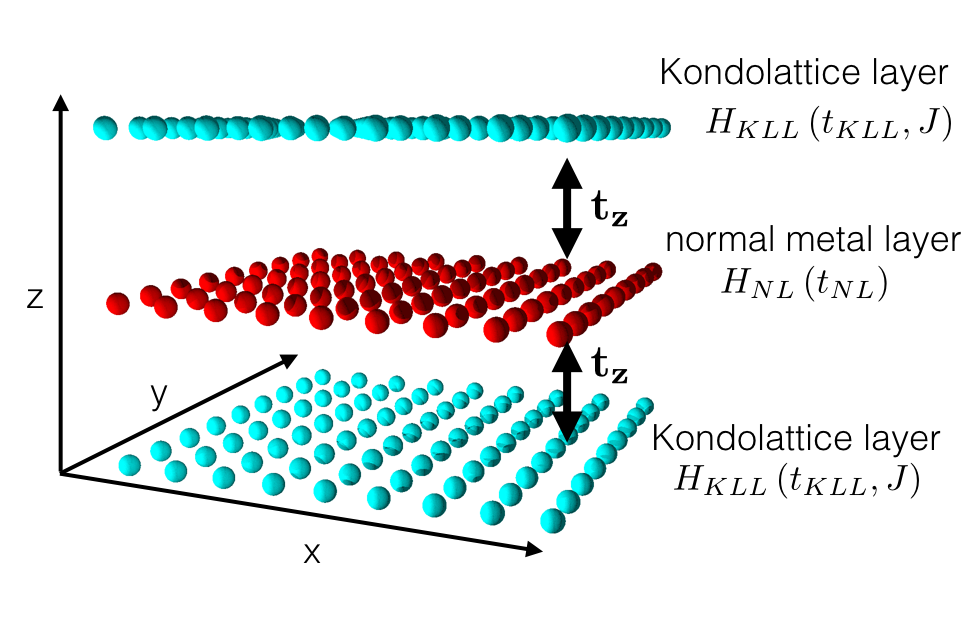}
\end{center}
\caption{Visualization of the Hamiltonian.
\label{Fig1}}
\end{figure}
Similar to our previous paper in which we analyze the Kondo effect in $f$ electron superlattices,\cite{PhysRevB.88.155134} we study a system consisting of a periodic structure made of Kondo lattice layers,\cite{doniach77,lacroix1979,fazekas1991}   described by $H_{KLL}$, and normal metallic layers, described by $H_{NL}$. We use the notation of a $(N,M)$ superlattice, where $N$ Kondo lattice layers are separated by $M$ normal metal layers, but will only focus on $(1,M)$ superlattices.
We model each layer as a square lattice. The model Hamiltonian, which is visualized in Fig. \ref{Fig1}, thus reads
\begin{eqnarray*}
H&=&H_{NL}+H_{KLL}+H_{inter},\\
H_{NL}&=&t_{NL}\sum_{<i,j>\sigma}c^\dagger_{iz\sigma}c_{jz\sigma},\\
H_{KLL}&=&t_{KLL}\sum_{<i,j>\sigma}c^\dagger_{iz\sigma}c_{jz\sigma}+J\sum_i\vec{S}_{iz}\cdot\vec{s}_{iz},\\
H_{inter}&=&t_{z}\sum_i\sum_{<z_1,z_2>\sigma}c^\dagger_{iz_1,\sigma}c_{iz_2\sigma}.
\end{eqnarray*}
The operator $c^\dagger_{iz\sigma}$ creates an electrons at lattice site $(i,z)$ with spin-direction $\sigma=(\uparrow,\downarrow)$, where $i$ is an index describing the $x$ and $y$ direction. $t_{NL}$ and $t_{KLL}$ are hopping constants within the normal metallic layers and the Kondo lattice layers, respectively. $t_z$ describes the hopping between different layers.
In our calculation we only study isotropic electron hopping, $t=t_{NL}=t_{KLL}=t_z$. $J$ describes the spin-spin interaction between the magnetic moments and the conduction electrons in the Kondo layer lattices. Throughout this paper we assume an antiferromagnetic coupling, $J>0$. Furthermore, we take the hopping $t$ as unit of energy.
All calculations in this study are performed for half filled lattices, $\langle n_\uparrow\rangle + \langle n_\downarrow\rangle=1$. The doped case is left for future studies.

We use the real-space dynamical mean field theory (RDMFT)\cite{georges1996} to calculate magnetic properties of this system.  RDMFT approximates the self-energy for each lattice site as local; the self-energy vanishes between different atoms, thus nonlocal fluctuations are not included into this framework. This approximation becomes exact in infinite dimensions. Because the self-energy can depend on the lattice site and spin direction, this approach is suitable to analyze the competition of the Kondo effect and magnetism in $f$ electron superlattices. 
In order to calculate a self-energy, each lattice site is mapped onto its own impurity model. This is done by calculating the local Green's function for each lattice site,
which can be written as 
\begin{displaymath}
G_{iz\sigma}(\omega+i\eta)=\frac{1}{\omega+i\eta-\Delta_{iz\sigma}(\omega+i\eta)-\Sigma_{iz\sigma}(\omega+i\eta)},
\end{displaymath}
where $\Sigma_{iz\sigma}(\omega+i\eta)$ is the self-energy for this lattice site.
The hybridization function $\Delta_{iz\sigma}(\omega+i\eta)$ describes the coupling between an impurity and conduction electrons. 
The resulting impurity models for each lattice site is then solved using the numerical renormalization group (NRG).\cite{wilson1975,bulla2008,peters2006,weichselbaum2007}

\section{Magnetization}
\label{sec_Mag}
As stated above, the properties of $f$ electron materials are strongly influenced by the competition between the RKKY interaction and the Kondo effect.
A similar competition will also occur in an $f$ electron superlattice. 
However, while in an ordinary $f$ electron material the strength of the RKKY interaction and the energy scale of the Kondo effect are fixed, in an $f$ electron superlattice they depend on the width of the layers used in the superlattice.

\begin{figure}[t]
\begin{center}
  \includegraphics[width=\linewidth]{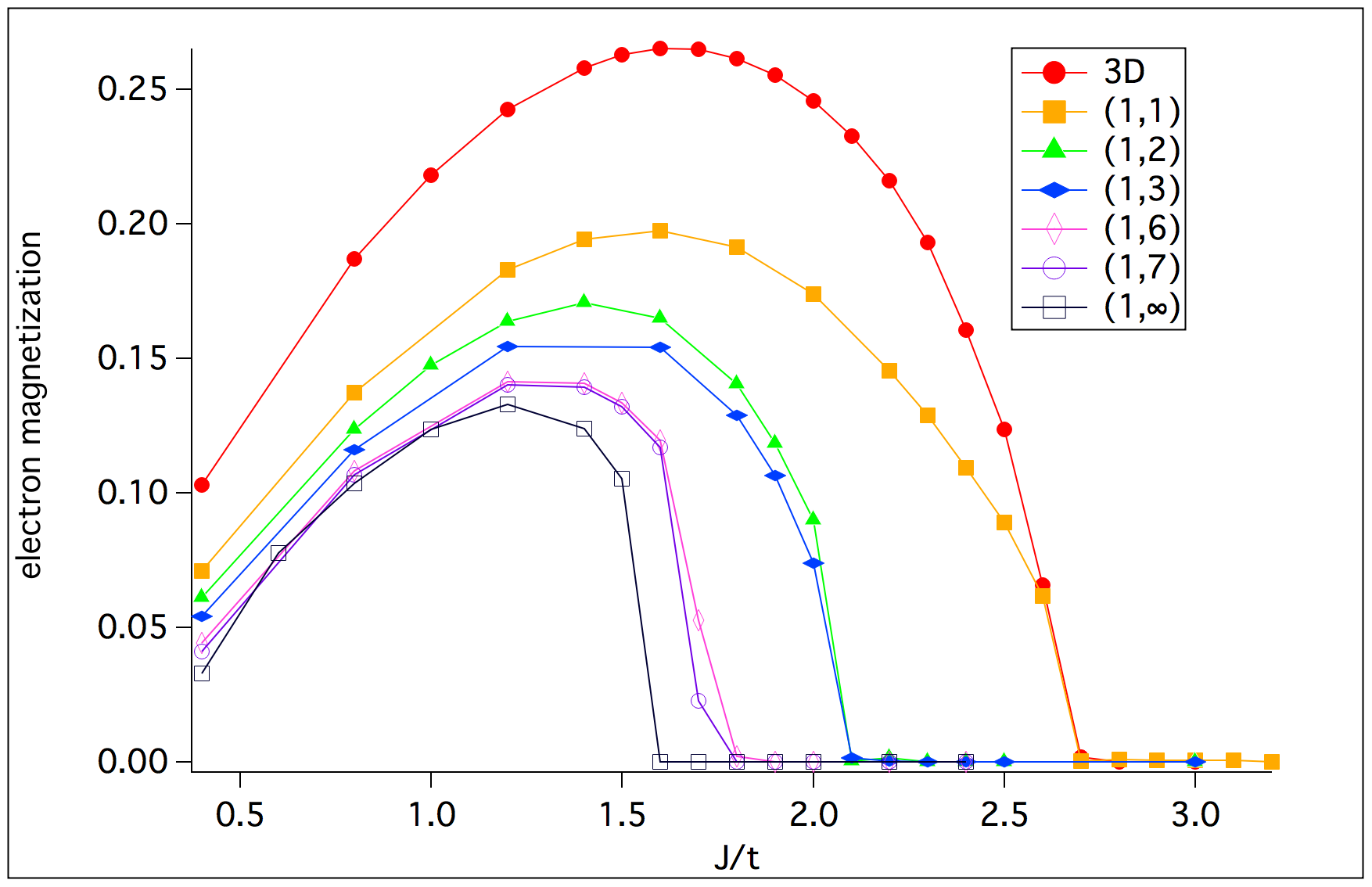}
\end{center}
\caption{Magnetization of the conduction electrons plotted against the coupling
  strength, $J$,  for different superlattices $(1,M)$. The magnetization curves
  interpolate between a single layer embedded in a 3D metallic host, $(1,\infty)$, and the 3D Kondo lattice model, $(1,0)$.
\label{Fig2}}
\end{figure}

Thus, we start our analysis by showing the magnetization of the conduction
electrons within the $f$-electron layers for different superlattice structures at $T=0$, see Fig. \ref{Fig2}. In the strong coupling region, $J/t>2.6$, all magnetization curves vanish independently of the superlattice. In this region all layers of the superlattice become paramagnetic, i.e. the localized spins are screened by the Kondo effect. In this phase the density of states vanishes exactly at the Fermi energy in all $f$ electron layers due to the formation of a Kondo insulating state, while the spacer layers stay metallic.

\begin{figure}[t]
\begin{center}
\includegraphics[width=\linewidth]{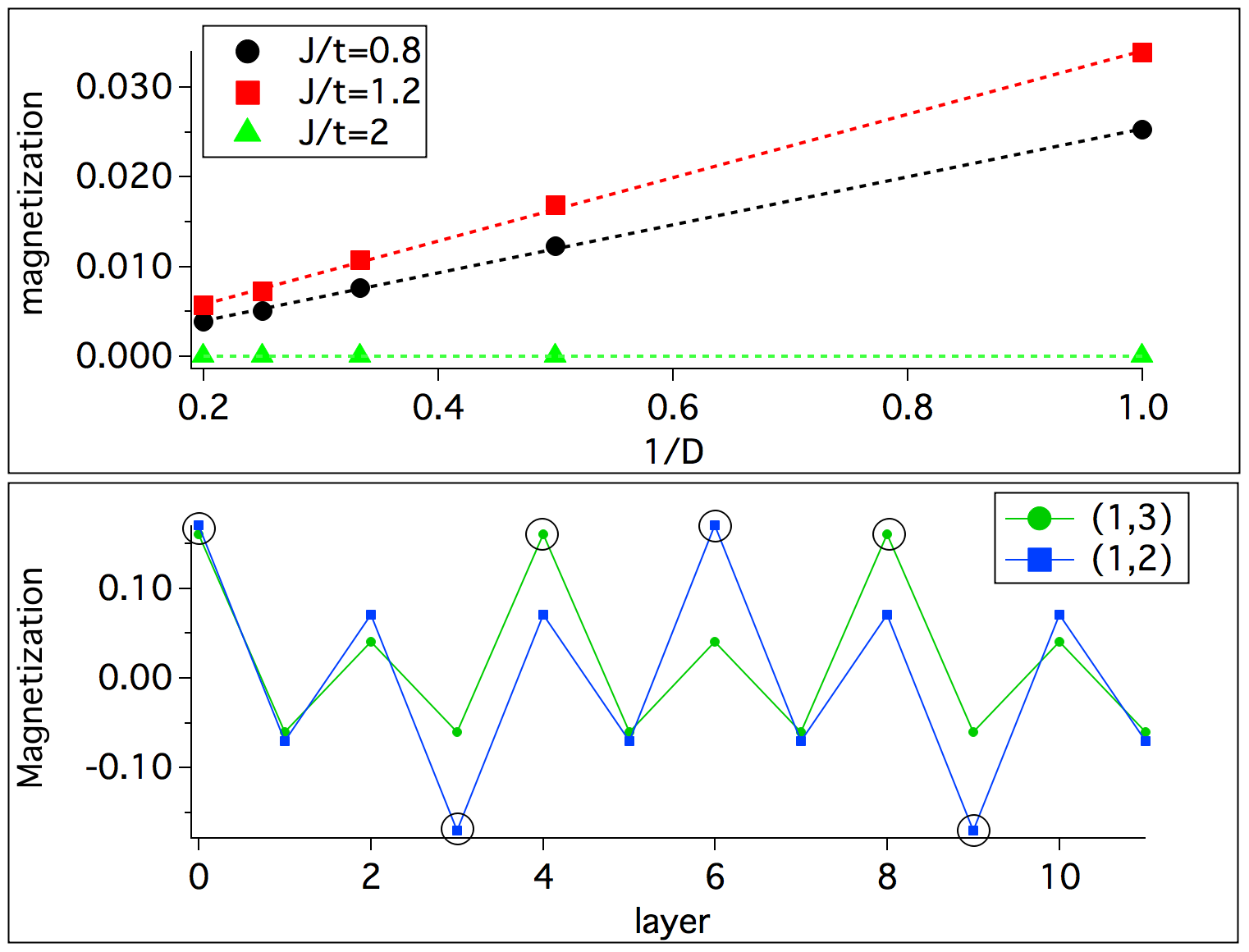}
\end{center}
\caption{Top: Local magnetization of the conduction electrons of an $f$ electron layer embedded into a 3D metal, $(1,\infty)$-superlattice, for different distances from the $f$ electron layer.  Bottom: Magnetization of the conduction electrons plotted against the layer for an $(1,3)$- and $(1,2)$-superlattices and $J/t=1.6$. We have marked the $f$ electron layers of both superlattices with circles. 
\label{Fig3}}
\end{figure}
The critical interaction strength, at which the magnetization vanishes, depends thereby on the superlattice structure.
The 3D Kondo lattice without any spacer layers, red curve in Fig. \ref{Fig2}, has the largest critical interaction strength, $J_C^{3D}/t\approx 2.6$.
We observe that the magnetization of the $(1,1)$ superlattice vanishes nearly at the same critical interaction strength as the 3D lattice.
Increasing the number of spacer layers further, the maximum value of the magnetization as well as the critical interaction strength decreases. 
Remarkably, we find that the $(1,2)$-, $(1,3)$-superlattices and $(1,4)$-, $(1,5)$-superlattices exhibit similar magnetization curves. 
The same behavior can also be observed for larger distances between $f$ electron layers; $(1,2n)$- and $(1,2n+1)$-superlattice exhibit similar magnetization curves. Thus, the quantum critical point in the $f$ electron superlattice changes stepwise. We will elucidate this point below after presenting more data.

We find that even a single $f$ electron layer, $(1,\infty)$ in Fig. \ref{Fig2}, which is embedded in a 3D metallic host, shows magnetism for $J/t<1.6$. This system corresponds to a superlattice with infinitely many spacer layers.

As mentioned before, DMFT does not include nonlocal fluctuations. It can be expected that the inclusion of nonlocal fluctuations will further reduce this critical value. However, a finite critical coupling strength, $J/t>0$, even for a single $f$-electron layer embedded in a 3D metallic host is thereby consistent with studies of (isolated) two-dimensonal Kondo lattices including nonlocal fluctuations,\cite{Fazekas1992,PhysRevB.87.205144} which indeed show a finite critical coupling strength.

Different superlattices interpolate between the 3D Kondo lattice model and a single $f$-electron layer within a 3D metallic host. 
We note that a two dimensional (2D) Kondo lattice, without any coupled metallic layers, exhibits a different magnetization curve with a larger maximal magnetization than the $(1,\infty)$ superlattice and cannot be directly compared to a superlattice.

The coupling between the $f$-electron layer and the metallic spacer
layers, induces a magnetization into the spacer
layers, which we show in Fig. \ref{Fig3}. 
Due to a Fermi surface nesting of $(\pi,\pi)$, the sign of the
magnetization oscillates from layer to layer. Thus, depending on the distance between different $f$ electron layers, these layers are coupled either ferromagnetically or antiferromagnetically, which has also been found in a weak coupling study of the
periodic Anderson model for
superlattices.\cite{PhysRevB.92.035129}

Naturally, the magnetization in a spacer layer decreases with increasing distance to the $f$ electron layer. The magnetism in the superlattice is purely due to the interaction of the localized spins and conduction electrons in the $f$ electron layer.  
We show the absolute value of the magnetization of the conduction electrons as a function of the distance of an $f$ electron layer in the top panel of Fig. \ref{Fig3}. We see that the induced magnetization behaves as $\vert n_\uparrow - n_\downarrow\vert\sim1/D$, where $D$ is the distance of the conduction electron to the $f$-electron layer. We note that these data could also be fitted by different power laws, because the DMFT-data only include short distances. However, an analysis of the electron susceptibility and a test calculation using $100$ layers, predict a behavior as $1/D$.
The prefactor of the $1/D$ law depends thereby on the coupling between the conduction electrons and the magnetic moments; for weak coupling the prefactor firstly increases, while for strong coupling the prefactor deceases and finally vanishes when the whole superlattice becomes nonmagnetic.
In the bottom panel of Fig. \ref{Fig3}, we show the magnetization of the conduction electrons in different layers in the $(1,2)$ and the $(1,3)$ superlattice. 
We have marked the $f$ electron layers in both superlattices by adding circles.
We observe that in the 
$(1,2)$ superlattice $f$ electron layers are coupled antiferromagnetically, while in the $(1,3)$ superlattice they are coupled ferromagnetically. 
As already mentioned above the magnetization curves of the $(1,2)$ and $(1,3)$ superlattice lie very close together. In Fig. \ref{Fig3}, we see that not only the magnetization in the $f$ electron layer is similar, but also in the next layer, see e.g. layer $1$.

\section{interlayer magnetic coupling}
\label{sec_coupling}
The induced magnetization leads to a magnetic coupling between different $f$-electron layers, which corresponds to the RKKY interaction between different magnetic layers. Such a magnetic coupling between different layers is well known from magnetic heterostructures.
While in a general case the magnetic interlayer coupling behaves as $1/D^2$, it has been shown that for a nested Fermi surface, like in our model, it behaves as $1/D$.\cite{PhysRevB.46.261}
In an $f$ electron superlattice, however, this magnetic interlayer coupling is modified by the Kondo effect. Such a competition is absent in ordinary magnetic heterostructures without $f$-electrons. We, therefore, analyze the influence of the Kondo effect on this interlayer magnetic coupling. 

To determine the interlayer coupling we prepare the following set-up:
\begin{enumerate}
\item We start with a fully converged magnetic solution of an $f$ electron superlattice. 
\item We select a single $f$-electron layer and set the self-energies of this layer to zero.
\item For all other layers, we use the self-energies of the converged solution.
\item We apply a staggered magnetic field to the selected $f$-electron layer, which points opposite to the original magnetization of the layer.
\end{enumerate}
We perform DMFT calculations for this set-up and study the magnetization of the selected layer as a function of the magnetic field. 
The self-energy of the selected layer is thereby updated during the DMFT iterations and calculated self-consistently.
In these calculations, there are two competing effects. On the one hand, due to the magnetic interlayer coupling, the $f$ electron layers of the superlattice without applied magnetic field try to restore the original solution. On the other hand, the magnetic field has a tendency to stabilize a solution where the magnetization of the selected layer is flipped. Thus, without an applied magnetic field the solution for this calculation will be identical to the input. Furthermore, if there is only the $f$ electron layer with applied magnetic field, an infinitesimally small magnetic field will be sufficient to stabilize a solution with flipped magnetization.
The strength of the magnetic field, at which the magnetization flips, corresponds to the magnetic coupling of the selected layer to all other $f$ electron layers in the superlattice.
\begin{figure}[t]
\begin{center}
  \includegraphics[width=\linewidth]{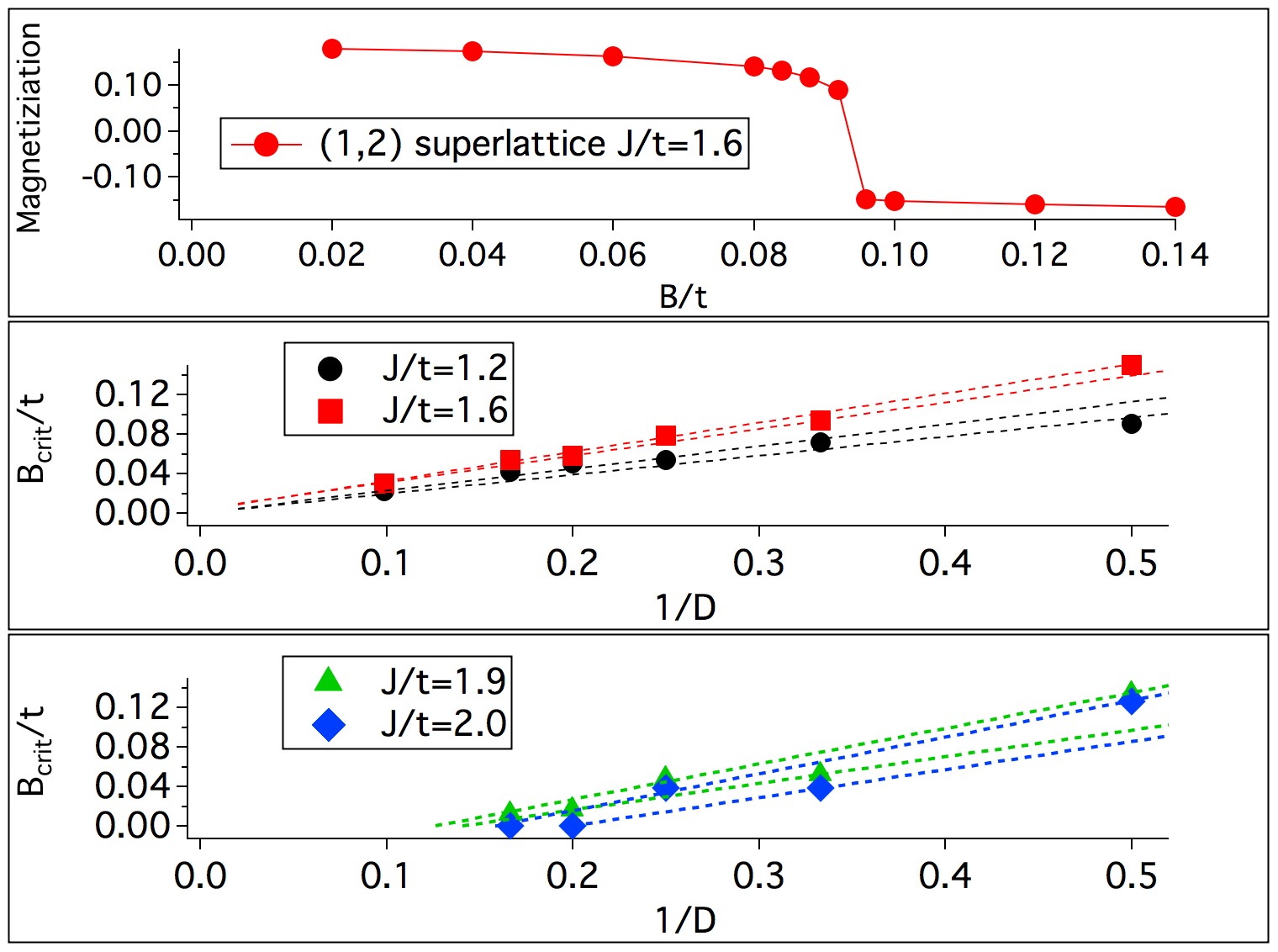}
\end{center}
\caption{Interlayer magnetic coupling calculated by applying a small magnetic field. The top panel shows the magnetization of the chosen layer as a function of the magnetic field for a $(1,2)$-superlattice with $J/t=1.6$. The middle and bottom panels show the interlayer coupling strength plotted against $1/D$ for weak coupling and strong coupling, respectively.
\label{Fig4}}
\end{figure}

We show the magnetic interlayer coupling, determined by the above procedure, in Fig. \ref{Fig4}. The top panel shows the magnetic-field dependence of the magnetization for an $(1,2)$-superlattice with coupling strength $J/t=1.6$. For weak magnetic field, we observe that the magnetization of the probed layer points into the same direction as in the initial solution. Increasing the strength of the magnetic field, the magnetization is reduced and flips at a critical strength of the magnetic field. This flipping of the magnetization is clearly visible as a jump. We take the value of this critical magnetic field strength as the magnetic interlayer coupling for this superlattice and interaction strength.

In the middle and bottom panel of Fig. \ref{Fig4}, we show this interlayer coupling plotted against $1/D$, where $D$ is the distance between the $f$ electron layers. Thus, the distance in a $(1,M)$-superlattice is $D=M+1$. We find that for weak coupling between the magnetic moments and the conduction electrons, $J/t=1.2$, the magnetic interlayer coupling can be well described as $1/D$. This agrees with the behavior of the magnetization shown in Fig. \ref{Fig3}.
Thus, there is a long-range interlayer coupling between different $f$ electron layers. Only for $D\rightarrow\infty$ the interlayer coupling vanishes completely.
For small coupling strength, $J$, the Kondo effect is exponentially weak and  does not play an important role. All the calculations have been performed for a half filled system so that the noninteracting Fermi surface is nested. Our results thus agree for weak coupling with the magnetic interlayer coupling of usual magnetic heterostructures.

\begin{figure}[t]
\begin{center}
  \includegraphics[width=\linewidth]{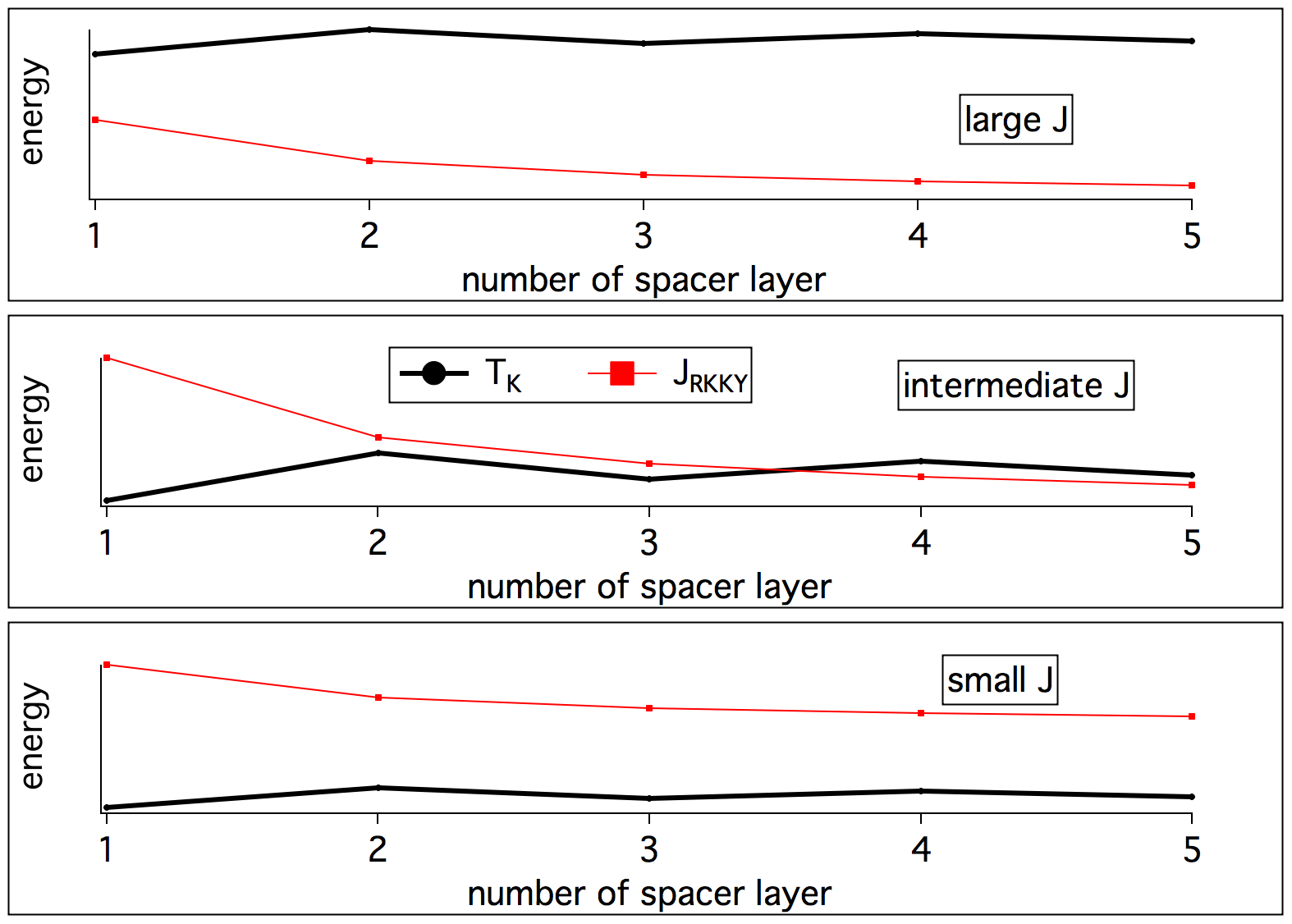}
\end{center}
\caption{Visualization of the competition between the Kondo effect and the RKKY interaction. We model the strength of the Kondo effect as an oscillating function in accordance with our previous study. The strength of the RKKY interaction decreases as $1/D$ with increasing distance. Top: large $J$, the Kondo effect is stronger than the RKKY for any superlattice. Middle panel: for intermediate $J$, the RKKY interaction is stronger than the Kondo effect for small $M$, but becomes weaker for large $M$. Bottom: For weak coupling $J$, the RKKY interaction is stronger than the Kondo effect for any superlattice. 
\label{Fig5}}
\end{figure}
For strong coupling, see bottom panel in Fig. \ref{Fig4}, the determined interlayer coupling cannot be fitted as $1/D$. The magnetic interlayer coupling  clearly deviates from a $1/D$ behavior and
vanishes already for $1/D>0$, where  the superlattice becomes nonmagnetic. 
This results from the competition between the RKKY interaction and the Kondo effect, and is thus a characteristic of $f$ electron superlattices. The Kondo effect screens the magnetic moments arising from the interacting $f$ electrons, and thus affects the magnetic interlayer interaction. 
A remarkable effect can be observed for strong coupling; the magnetic interlayer coupling for $1/D=0.2$ ($M=3$) and $1/D=0.25$ ($M=2$) are nearly identical. The interlayer coupling vanishes pairwise for increasing number of spacer layer. We have thus included separate least square fits (dashed lines) for even number and odd number of spacer layers into Fig. \ref{Fig4}. Even for weak coupling we see that superlattices with even and odd number of spacer layers, have slightly different least square fits. The difference between these two lines corresponds qualitatively to the strength of the Kondo effect in the superlattice.

The reason for this even-odd effect, i.e. pairwise vanishing of the interlayer coupling, is that the Kondo temperature and the RKKY interaction change, but both effects cancel each other. Not only the RKKY interaction, but also the Kondo effect and the Kondo temperature depend on the superlattice.\cite{PhysRevB.88.155134} The Kondo temperature shows even-odd oscillations in $f$ electron superlattices depending on whether the number of spacer layers is even or odd. 
This effect becomes important when analyzing the competition between the RKKY interaction and the Kondo effect.
The Kondo effect becomes weaker when changing the number of spacer layers from even to odd. Thus, the magnetic moments are less screened, which effectively increases the magnetic interlayer coupling. On the other hand, increasing the distance between the $f$ electron layers leads to a decrease of the interlayer coupling. 
As a result, the observed magnetic interlayer coupling, which takes into account both RKKY and Kondo effect, does not change.
This explains the stepwise decrease of the interlayer coupling and the existence of two different superlattices with similar magnetization curves in Fig. \ref{Fig2}.

Taking the results of the magnetic interlayer coupling, we can now understand the competition between RKKY interaction and the Kondo effect in $f$ electron superlattices. 
We can distinguish different situations, which we qualitatively show in Fig. \ref{Fig5}.
The interlayer RKKY interaction behaves as $1/D$ in the superlattice. Furthermore, there is a magnetic intralayer coupling between the localized moments within the same layer, which does not depend on the structure of the superlattice.  
For weak coupling, see bottom panel in Fig. \ref{Fig5}, the Kondo temperature is exponentially small and even the magnetic intralayer coupling is stronger than the Kondo effect. In this situation, we find a magnetic ground state independent of the number of spacer layers. Furthermore, because the Kondo effect is negligible, we observe the $1/D$ dependence of the interlayer coupling as for usual magnetic heterostructures. For strong coupling between magnetic moments and conduction electrons, as shown in the top panel in Fig. \ref{Fig5}, the Kondo effect is stronger than the RKKY interaction in any superlattice. In this situations we can only find paramagnetic ground states. There is no magnetic interlayer coupling, because all magnetic moments are completely screened by the Kondo effect. 
The most interesting case occurs when the intralayer RKKY coupling is smaller than the Kondo temperature, but  the sum of intralayer and interlayer RKKY interaction is larger, see middle panel of Fig. \ref{Fig5}. 
Increasing the number of metallic spacer layers leads then to a reduction of the RKKY interaction, which finally becomes weaker than the Kondo effect and thereby leads to a vanishing of the magnetic order. For this intermediate coupling strength, the Kondo effect and the RKKY interaction strongly influence each other. The Kondo effect becomes slightly weaker when increasing the number of spacer layers from even to odd. However, because at the same time also the strength of the RKKY interaction decreases, both effects may cancel out each other. 

\section{Spectral functions}
\label{sec_spectral}
\begin{figure}
\begin{center}
\includegraphics[width=0.9\linewidth]{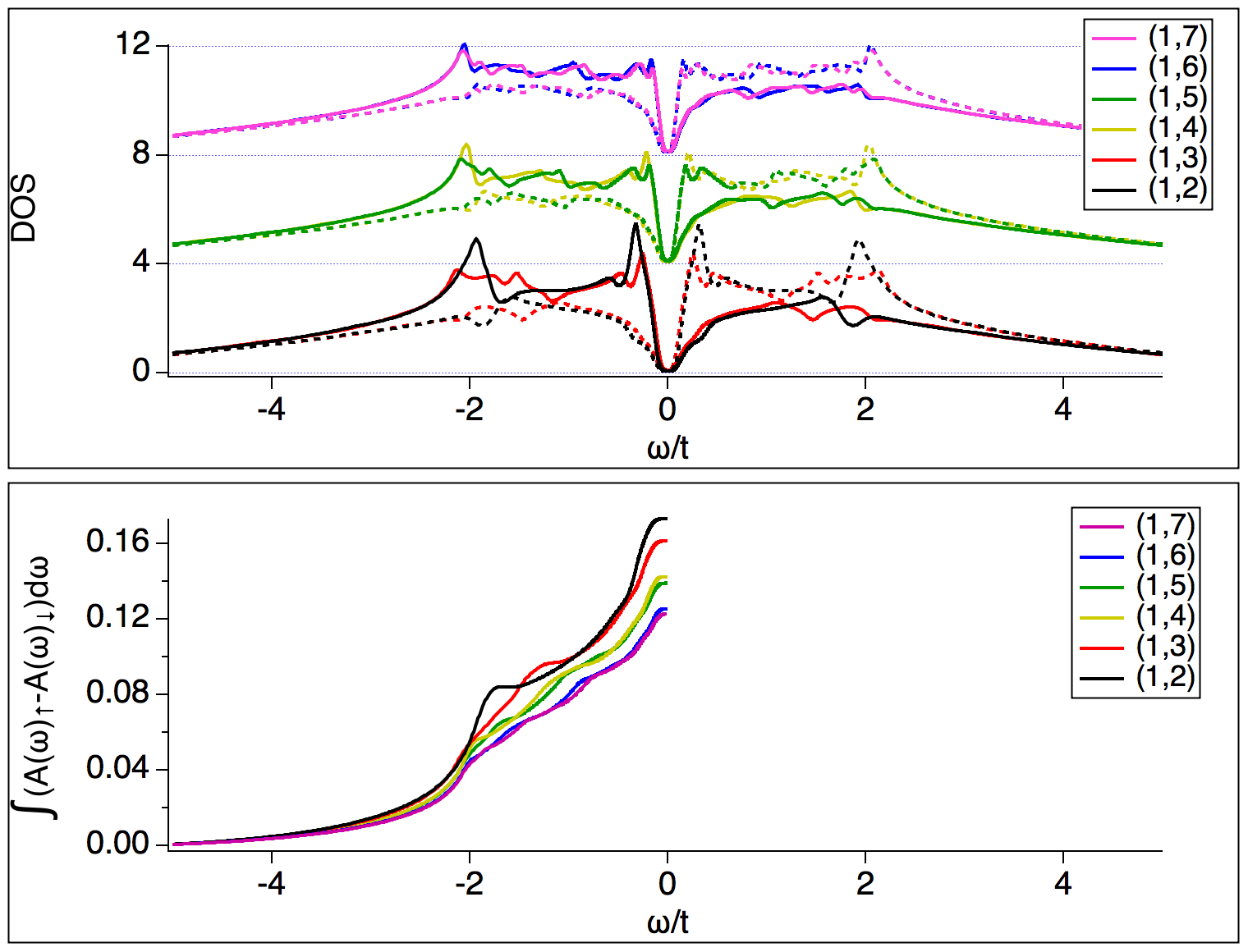}
\includegraphics[width=0.9\linewidth]{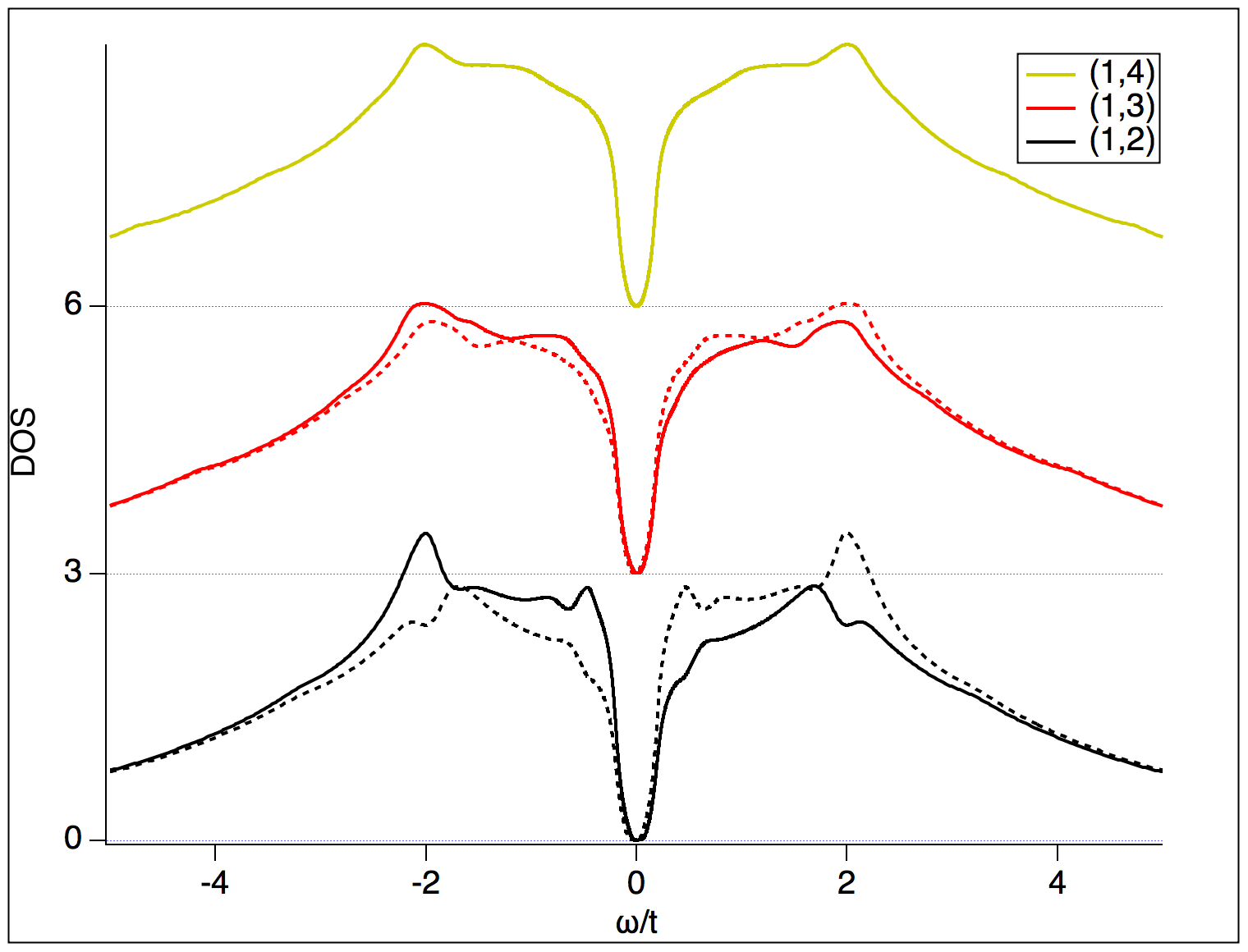}
\end{center}
\caption{Spectral functions of the conduction electrons in the $f$ electron layer in the ordered phase. Solid lines and dashed lines correspond to the majority and minority spin direction on the analyzed lattice site, respectively. Top: Spectral functions for six different superlattices and $J/t=1.6$. (We have shifted the origin of the spectral functions of the upper curves for clarity) We always show the spectral function of two superlattices with similar magnetization together. Middle: Integrated spectral weight $\int (A_\uparrow-A_\downarrow)$ for the spectral functions in the top panel. Bottom: Spectral functions for different superlattices and $J/t=2$. For $M>3$ the magnetization of the superlattice vanishes.
\label{Fig6}}
\end{figure}
Finally, we discuss the spectral functions in the ordered phase, shown in Fig. \ref{Fig6}. In the top panel, we show the spectral functions of the conduction electrons in the $f$ electron layer for $J/t=1.6$ and different superlattices. At this interaction strength, the magnetic order persists for the shown superlattices and
the spectral functions of the conduction electrons form a gap at the Fermi energy, $\omega=0$. This gap is partly due to the magnetic order and partly due to the Kondo effect which appears in this layer. Due to the absence of $f$ electrons in the spacer layer, the spectral functions in the spacer layer remain metallic although they are magnetically ordered. Besides the gap, the most prominent features in the spectral functions are a van-Hove singularity close to the gap and an excitation at $\omega=2t$. The strength of the van-Hove singularity decreases with increasing number of spacer layers $M$, which might be related to the decrease of the magnetization.
In the middle panel we show $\int (A_\uparrow (\omega)-A_\downarrow (\omega))d\omega$, which corresponds to the magnetization, for the superlattices shown in the top panel. We observe again the stepwise change of the magnetization when increasing the number of spacer layers, i.e. for $M>3$ this magnetizations  are nearly equal for superlattices $(1,4)$ and $(1,5)$ as well as for $(1,6)$ and $(1,7)$. In the top panel we demonstrate that not only the magnetization is similar in these superlattices, but also the spectral function.  Excitations lie approximately at the same energies and have the same strength. 
In the bottom panel of Fig. \ref{Fig6}, we show the spectral functions for $J/t=2$. The magnetization vanishes when increasing the number of spacer layers $M$, although the coupling between localized moments and conductions electrons is constant.  

\section{Conclusions}
We have analyzed the magnetic order in $f$ electron superlattices and have demonstrated that the quantum critical point of the $f$ electron material can be tuned by changing the superlattice structure similar to experimental results on the CeIn$_3(n)$/LaIn$_3$(4) superlattices\cite{Shishido2010}. We have focused in this study on the influence of the superlattice on the competition between the RKKY interaction and the Kondo effect. Besides the RKKY interaction and the Kondo effect also the reduced dimensionality of the $f$-electron material will become important, especially when the number of $f$-electron layers becomes small as in the experiments on CeIn$_3(n)$/LaIn$_3(4)$-superlattices. Increased magnetic fluctuations due to the reduced dimensionality can further reduce the critical coupling of the magnetic order. An analysis of the interplay of RKKY interaction, Kondo effect, and nonlocal fluctuations is left for a future study using cluster extensions of the dynamical mean field theory.

Furthermore, we have studied the competition of the Kondo effect and the RKKY interaction in superlattices.
We have demonstrated that while the magnetic interlayer coupling between different $f$ electron layers behaves as $1/D$ for weak Kondo effect, 
as in non-$f$-electron superlattices, it differs strongly from this behavior for strong Kondo effect.  In this case the magnetic interlayer coupling vanishes already for a finite number of spacer layers. Furthermore, due to the interplay between the Kondo effect and the RKKY interaction, the interlayer coupling vanishes stepwise when increasing the number of spacer layers.  As a consequence, two superlattices with different number of spacer layers show similar magnetization curves.
\paragraph*{Acknowledgments ~}
This work was partly supported by a Grant-in-Aid for Scientific Research 
on Innovative Areas (JSPS KAKENHI Grant No. 15H05855), JSPS 
Grant-in-Aid for Scientific Research  (No. 16K05501, No. 26800177), and
Grant-in-Aid for Program for Advancing Strategic International Networks to
Accelerate
the Circulation of Talented Researchers (Grant No. R2604).

\bibliography{paper}

\end{document}